\documentclass{jpsj-suppl}
\usepackage{txfonts} 
\usepackage{subfigure}

\title{Search for R-parity violation from J-PARC and LHC}

\author{Masato \textsc{yamanaka}$^{1}$}

\inst{$^{1}$Department of Physics, Nagoya University, Nagoya 464-8602, Japan}

\email{yamanaka@eken.phys.nagoya-u.ac.jp}

\recdate{Sep. 7, 2014}

\abst{We consider the case that $\mu$-$e$ conversion 
signal is discovered but other charged lepton flavor violating 
(cLFV) processes will never be found. In such a case, we need 
other approaches to confirm the $\mu$-$e$ conversion and 
its underlying physics without conventional cLFV searches. 
We study R-parity violating (RPV) SUSY models as a benchmark. 
We briefly review that our interesting case is realized in 
RPV SUSY models with reasonable settings according to current 
theoretical/experimental status. 
We focus on the exotic collider signatures at the LHC ($pp \to 
\mu^- e^+$ and $pp \to jj$) as the other approaches. We 
show the correlations between the branching ratio of $\mu$-$e$ 
conversion process and cross sections of these processes. 
It is first time that the correlations are graphically shown. 
We exhibit the RPV parameter dependence of the branching 
ratio and the cross sections, and discuss the feasibility to 
determine the parameters. 
This paper is based on Ref.~\cite{RPV}. }

\kword{$\mu$-$e$ conversion, charged lepton flavor violation, 
R-parity violating SUSY models, J-PARC, LHC}

\begin{document}
\maketitle

\section{Introduction}
\vspace{-2mm}

Lepton flavor violation (LFV) is the clearest signal for physics beyond the Standard 
Model (SM)~\cite{Kuno:1999jp}, and extensive searches for LFV have been made 
since the muon was found~\cite{Brooks:1999pu, Adam:2013mnn, Bertl:2006up, 
Bellgardt:1987du}.  
Though a lot of efforts have been made, we have not found any LFV signals with
charged leptons.  LFV had, however, been found in neutrino oscillation
\cite{Fukuda:1998mi,Abe:2013hdq} and it indeed requires us to extend the
SM so that physics beyond the SM must include LFV.  This fact also gives
us a strong motivation to search for charged lepton flavor violation (cLFV).

Along this line new experiments to search for cLFV will start soon. COMET 
\cite{Cui:2009zz,Kuno:2013mha} and DeeMe \cite{Natori:2014yba} 
will launch within a few years and search $\mu$-$e$ conversion.  In these 
experiments, first, muons are trapped by target nucleus, then, if cLFV 
exists, it converts into an electron.
If COMET/DeeMe observe the $\mu$-$e$ conversion, then with what
kind of new physics should we interpret it?  Now it is worth 
considering since we are in-between two kinds of cLFV experiments
with muon.

For these several decades, theories with supersymmetric extension have
been most studied. These theories include a source of LFV.  It is
realized by the fact that the scalar partner of charged leptons have a 
different flavor basis from that of charged leptons.  In
addition, R-parity is often imposed on this class of the
theory\cite{Hisano:1995cp,Sato:2000ff}. With it, $\mu\rightarrow
e\gamma$ process has the largest branching ratio among the three cLFV
processes. This occurs through the dipole process depicted and the 
other two, $\mu-e$ conversion and $\mu\rightarrow 3e$, are 
realized by attaching a quark line and an
electron line at the end of the photon line respectively, giving an
O($\alpha$) suppression.  Those branching ratios must be smaller than
that of $\mu\rightarrow e\gamma$.  At this moment, however, the upper
bounds for those branching ratios are almost same each other. It means
if COMET/DeeMe observe the $\mu-e$ conversion, we have to discard 
this scenario.

It is, however, possible to find a theory easily in which COMET/DeeMe 
find cLFV first. 
To see this we first note that the $\mu\rightarrow e \gamma$ process 
occurs only at loop level due to the gauge invariance,
while other two can occur as a tree process. Therefore in this case we
have to consider a theory in which the $\mu-e$ conversion process occurs
as tree process.
In other words we have to assume a particle which violate
muon and electron number. Since $\mu - e$ conversion occurs in a nucleus,
it also couples with quarks with flavor conservation. 
Furthermore it is better to assume that it does not couple with two electrons 
as we have not observed $\mu\rightarrow 3e$.

In this paper we consider the case that COMET/DeeMe indeed observe 
the cLFV process, while all the other experiments will not observe 
anything new at that time.  With this situation, we need to understand 
how to confirm the cLFV in other experiments. 
Unfortunately in this case other new physics signals are expected to be
quite few, since the magnitude of the cLFV interaction is so small due
to its tiny branching ratio. Therefore it is very important to simulate
now how to confirm the COMET signal and the new physics.  As a benchmark
case we study supersymmetric models without R
parity~\cite{deGouvea:2000cf} .  In this kind of theory the scalar lepton
mediates $\mu \leftrightarrow e$ flavor violation.

\section{RPV interaction and our scenario}  \label{Sec:interaction} 
\vspace{-2mm}

In general the supersymmetric gauge invariant superpotential contains 
the R-parity violating terms~\cite{Weinberg:1981wj, Sakai:1981pk, 
Hall:1983id}, 
\begin{equation}
\begin{split}
   \mathcal{W}_\text{RPV} = \lambda_{ijk} L_i L_j E_k^c 
   + \lambda'_{ijk} L_i Q_j D_k^c 
   + \lambda''_{ijk} U_i^c D_j^c D_k^c, 
\label{Eq:RPV_SP}   
\end{split}      
\end{equation}
where $E_i^c$, $U_i^c$ and $D_i^c$ are $SU(2)_L$ singlet superfields, 
and $L_i$ and $Q_i$ are $SU(2)_L$ doublet superfields. Indices $i$, $j$, 
and $k$ represent the generations. 
We take $\lambda_{ijk} = - \lambda_{jik}$ and $\lambda''_{ijk} 
= - \lambda''_{ikj}$. First two terms include lepton number violation, and 
the last term includes baryon number violation. Since some combinations 
of them accelerate proton decay, we omit the last term.

Our interesting situation is that only $\mu$-$e$ conversion is discovered,
and other cLFV processes will never be observed. The situation is realized 
under the following 3 setting on the RPV interaction: 
\begin{enumerate}
\item 
only the third generation slepton contributes to the RPV interactions 
\item
for quarks, flavor diagonal components are much larger than that 
of off-diagonal components, i.e., CKM-like matrix, $\lambda'_{ijj} 
\gg \lambda'_{ijk} (j \neq k)$
\item 
the generation between left-handed and right-handed leptons are 
different, $\lambda_{ijk} (i \neq k \text{ and } j \neq k)$. 
\end{enumerate}
The setting-1 is naturally realized by the RG evolved SUSY spectrum with 
universal soft masses at the GUT scale. For the simplicity, we decouple 
SUSY particles except for the third generation sleptons. 
The setting-2 is also realized in most cases unless we introduce additional 
sources of flavor violations. 
The setting-3 is artificially introduced to realize the interesting situation 
(see Introduction).
Under the settings, the Lagrangian from the superpotential~\eqref{Eq:RPV_SP}
is reduced as follows, 
\begin{equation}
\begin{split}
   &
   \mathcal{L}_\text{RPV} 
   = \mathcal{L}_{\lambda} 
   + \mathcal{L}_{\lambda'}, 
   \\&
   \mathcal{L}_\lambda = 2 \bigl[ 
   \lambda_{312} \tilde \nu_{\tau L} \overline{\mu} P_L e 
   + \lambda_{321} \tilde \nu_{\tau L} \overline{e} P_L \mu 
   + \lambda_{132} \tilde \tau_L \overline{\mu} P_L \nu_e 
   + \lambda_{231} \tilde \tau_L \overline{e} P_L \nu_\mu
   \\& \hspace{10mm}  
   + \lambda_{123} \tilde \tau_R^* \overline{(\nu_{eL})^c} P_L \mu 
   + \lambda_{213} \tilde \tau_R^* \overline{(\nu_{\mu L})^c} P_L e
   \bigr] + \text{h.c.}, 
   \\& 
   \mathcal{L}_{\lambda'} 
   = \bigl[ 
   \lambda'_{311} \bigl( \tilde \nu_{\tau L} \overline{d} P_L d 
   - \tilde \tau_L \overline{d} P_L u \bigr) 
   + \lambda'_{322} \bigl( \tilde \nu_{\tau L} \overline{s} P_L s 
   - \tilde \tau_L \overline{s} P_L c \bigr)
   \bigr] + \text{h.c.}.  
\label{Eq:RPV_L2}   
\end{split}      
\end{equation}

Some kind of processes described by the Lagrangian \eqref{Eq:RPV_L2} 
strongly depend on the values of $\lambda'_{311}$ and $\lambda'_{322}$. 
In order to clarify the dependence and to discuss the discrimination of each 
other, we study following three cases in our paper~\cite{RPV}: [case-I] 
$\lambda'_{311} \neq 0$ and $\lambda'_{322} = 0$, 
[case-I\hspace{-1pt}I] $\lambda'_{311} = 0$ and $\lambda'_{322} \neq 0$, 
and [case-I\hspace{-1pt}I\hspace{-1pt}I]  $\lambda'_{311} \neq 0$ 
and $\lambda'_{322} \neq 0$. 
In this talk, we focus on the case-I.

\section{Exotic processes in our scenario}  \label{Sec:Obs} 

In the scenario we have five types of exotic processes: 
\\[2mm]
~~~~~ 1 ~~ $\mu$-$e$ conversion in a nucleus ($\mu^- N \to e^- N$)
\\
~~~~~ 2 ~~ $\mu^- e^+$ production at LHC ($pp \to \mu^- e^+$) 
\\
~~~~~ 3 ~~ dijet production at LHC ($pp \to jj$)
\\
~~~~~ 4 ~~ non-standard interaction (NSI) of neutrinos
\\
~~~~~ 5 ~~ muonium conversion ($\mu^+e^- \to \mu^-e^+$)
\\[2mm]
In the situation that the $\mu$-$e$ conversion is discovered while 
other cLFV signals will never be found, we discuss the possibility whether 
we can confirm the $\mu$-$e$ conversion signal with the five types 
processes or not. 
Details of each process and the formulation of their reaction rates are 
given in our paper~\cite{RPV}.

Note that in our scenario other muon cLFV processes 
($\mu \to e \gamma$, $\mu \to 3e$, $\mu^- e^- \to e^- e^-$ 
in muonic atom~\cite{Koike:2010xr}, and so on) occur at two-loop level.  
At one glance the tau sneutrino can connect with the photon via 
d-quark loop. 
The contribution of the loop of the diagram is
\begin{eqnarray}
 \lambda'\left(-\frac{1}{3}\right) e \frac{m_d q_\mu}{8\pi^2}
\int_0^1 dx(1-2x)\log(m_d^2-(x-x^2)q^2)\ \propto q^2q^\mu,
\end{eqnarray}
where $q$ is the momentum of the photon. The contribution to cLFV
is, therefore vanish with
 on-shell photon ($q^2=0$) for $\mu\rightarrow e\gamma$
and with $\bar e \gamma_\mu e$ attached for $\mu\rightarrow 3e$
due to gauge symmetry($q^\mu\bar e\gamma_\mu e=0$).
Thus these processes occur at two-loop level. Furthermore these  
processes are extremely suppressed by higher order couplings, 
gauge invariance, and so on. Therefore we do not study these 
processes.

\section{Numerical result}  \label{Sec:result} 

\vspace{-2mm}
\begin{table}[t]
\caption{Current and future experimental limits on the $\mu$-$e$ 
conversion branching ratio 
and the upper limits on $\lambda' \lambda$ corresponding to each 
experimental limit. }
\vspace{-5mm}
\hspace{-3mm}
\small{
{\renewcommand\arraystretch{1.7}
\begin{tabular}{llllll}
\hline
Experiment 
& BR limit
& Limit on $\lambda'_{311} \lambda$ (case-I)
& Limit on $\lambda'_{322} \lambda$ (case-I\hspace{-1pt}I)
& Limit on $\lambda' \lambda$ (case-I\hspace{-1pt}I\hspace{-1pt}I)
\\ \hline \hline
SINDRUM
& $7 \times 10^{-13}$~\cite{Bertl:2006up}
& $1.633 \times 10^{-7} \Bigl( \dfrac{m_{\tilde \nu_\tau}}{1\text{TeV}} \Bigr)^2$
& $3.170 \times 10^{-7} \Bigl( \dfrac{m_{\tilde \nu_\tau}}{1\text{TeV}} \Bigr)^2$
& $1.072 \times 10^{-7} \Bigl( \dfrac{m_{\tilde \nu_\tau}}{1\text{TeV}} \Bigr)^2$
\\[0.5mm]
DeeMe
& $5 \times 10^{-15}$~\cite{Natori:2014yba}
& $ 1.550 \times 10^{-8} \Bigl( \dfrac{m_{\tilde \nu_\tau}}{1\text{TeV}} \Bigr)^2$
& $ 2.915 \times 10^{-8} \Bigl( \dfrac{m_{\tilde \nu_\tau}}{1\text{TeV}} \Bigr)^2$
& $ 1.012 \times 10^{-8} \Bigl( \dfrac{m_{\tilde \nu_\tau}}{1\text{TeV}} \Bigr)^2$
\\[0.5mm]
COMET-I 
& $7 \times 10^{-15}$~\cite{Kuno:2013mha}
& $1.830 \times 10^{-8} \Bigl( \dfrac{m_{\tilde \nu_\tau}}{1\text{TeV}} \Bigr)^2$ 
& $3.504 \times 10^{-8} \Bigl( \dfrac{m_{\tilde \nu_\tau}}{1\text{TeV}} \Bigr)^2$
& $1.196 \times 10^{-8} \Bigl( \dfrac{m_{\tilde \nu_\tau}}{1\text{TeV}} \Bigr)^2$
\\[0.5mm]
COMET-I\hspace{-1pt}I 
&$3 \times 10^{-17}$~\cite{Kuno:2013mha}
& $1.198 \times 10^{-9} \Bigl( \dfrac{m_{\tilde \nu_\tau}}{1\text{TeV}} \Bigr)^2$
& $2.294 \times 10^{-9} \Bigl( \dfrac{m_{\tilde \nu_\tau}}{1\text{TeV}} \Bigr)^2$
& $7.827 \times 10^{-10} \Bigl( \dfrac{m_{\tilde \nu_\tau}}{1\text{TeV}} \Bigr)^2$
\\[0.5mm]
PRISM 
& $7 \times 10^{-19}$~\cite{Kuno:2013mha}
& $1.830 \times 10^{-10} \Bigl( \dfrac{m_{\tilde \nu_\tau}}{1\text{TeV}} \Bigr)^2$
& $3.504 \times 10^{-10} \Bigl( \dfrac{m_{\tilde \nu_\tau}}{1\text{TeV}} \Bigr)^2$
& $1.196 \times 10^{-10} \Bigl( \dfrac{m_{\tilde \nu_\tau}}{1\text{TeV}} \Bigr)^2$
\\ \hline
\end{tabular} 
}
}
\label{Tab:mue_conv}
\end{table}

We are now in a position to show numerical results. 
Table~\ref{Tab:mue_conv} shows the current experimental limit and 
the future single event sensitivity for $\mu$-$e$ conversion process, 
and shows the upper limits on the combination of the RPV couplings, 
$\lambda' \lambda$, corresponding to the limit and the sensitivities in 
each experiment. 
In the calculation of the upper limits, we take Au, Si, and Al for target 
nucleus of SINDRUM-I\hspace{-1pt}I, DeeMe, and other experiments, 
respectively.

$\mu$-$e$ conversion search is a reliable probe to both the RPV 
couplings and tau sneutrino mass. 
The current experimental limit puts strict limit on the RPV couplings, 
$\lambda' \lambda \lesssim 10^{-7}$ for $m_{\tilde \nu_\tau} = 
1\text{TeV}$ and $\lambda' \lambda \lesssim 10^{-5}$ for 
$m_{\tilde \nu_\tau} = 3\text{TeV}$, respectively. 
In near future, the accessible RPV couplings will be extended by more 
than 3 orders of current limits, $\lambda' \lambda \simeq 10^{-10}$ 
for $m_{\tilde \nu_\tau} = 1\text{TeV}$ and $\lambda' \lambda 
\simeq 10^{-8}$ for $m_{\tilde \nu_\tau} = 3\text{TeV}$, 
respectively.

The $\mu$-$e$ conversion process is one of the clear signatures for 
the RPV scenario, but it is not the sufficient evidence of the scenario. 
We must check the correlations among the reaction rates of 
$\mu$-$e$ conversion process, the cross sections of $pp \to 
\mu^- e^+$ and $pp \to jj$, and so on in order to discriminate 
the case-I, -I\hspace{-1pt}I, and -I\hspace{-1pt}I\hspace{-1pt}I 
each other and to confirm the RPV scenario.

\begin{figure}[t!]
\hspace{-7mm}
\begin{tabular}{cc}
\subfigure[$m_{\tilde \nu_\tau} = 1$TeV. $\sqrt{s}= 14$TeV. ]{
\includegraphics[scale=0.61]{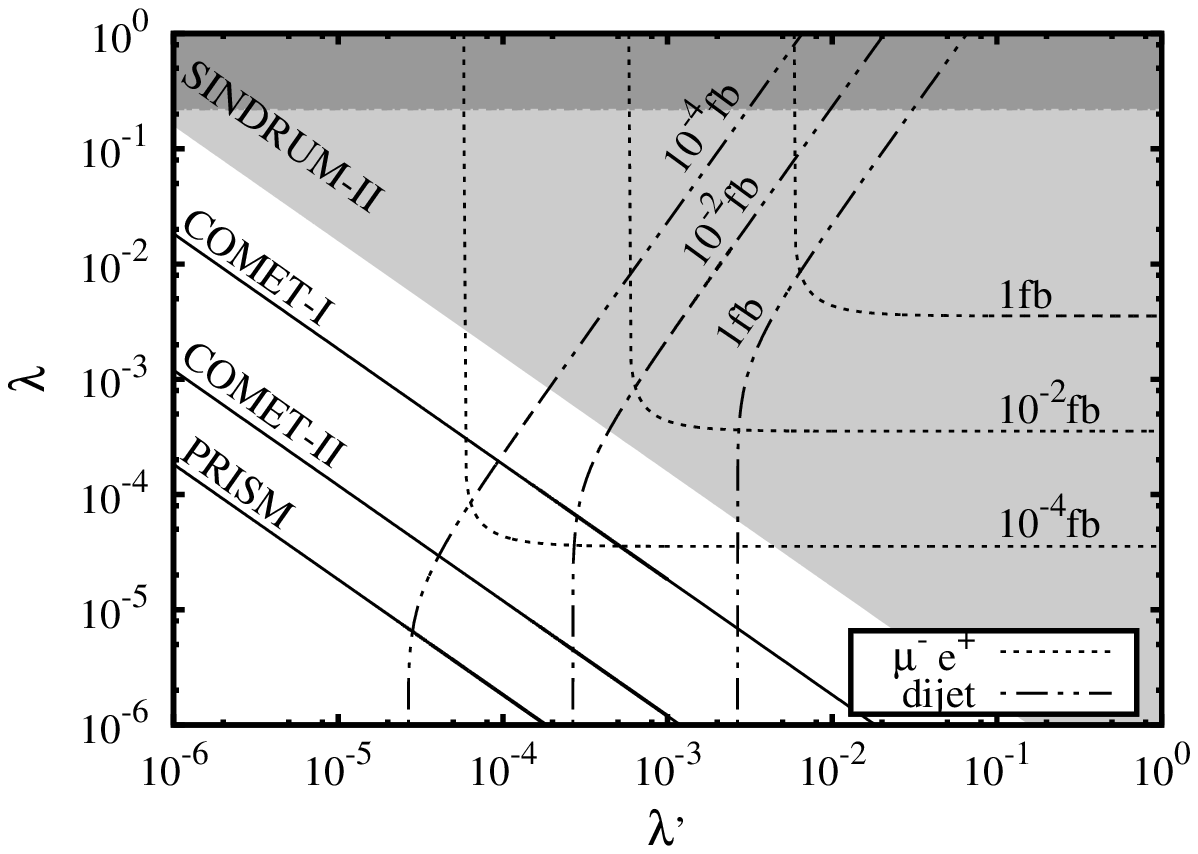}
\label{left1}
} & \hspace{-10mm}
\subfigure[$m_{\tilde \nu_\tau} = 1$TeV. $\sqrt{s}= 100$TeV. ]{
\includegraphics[scale=0.61]{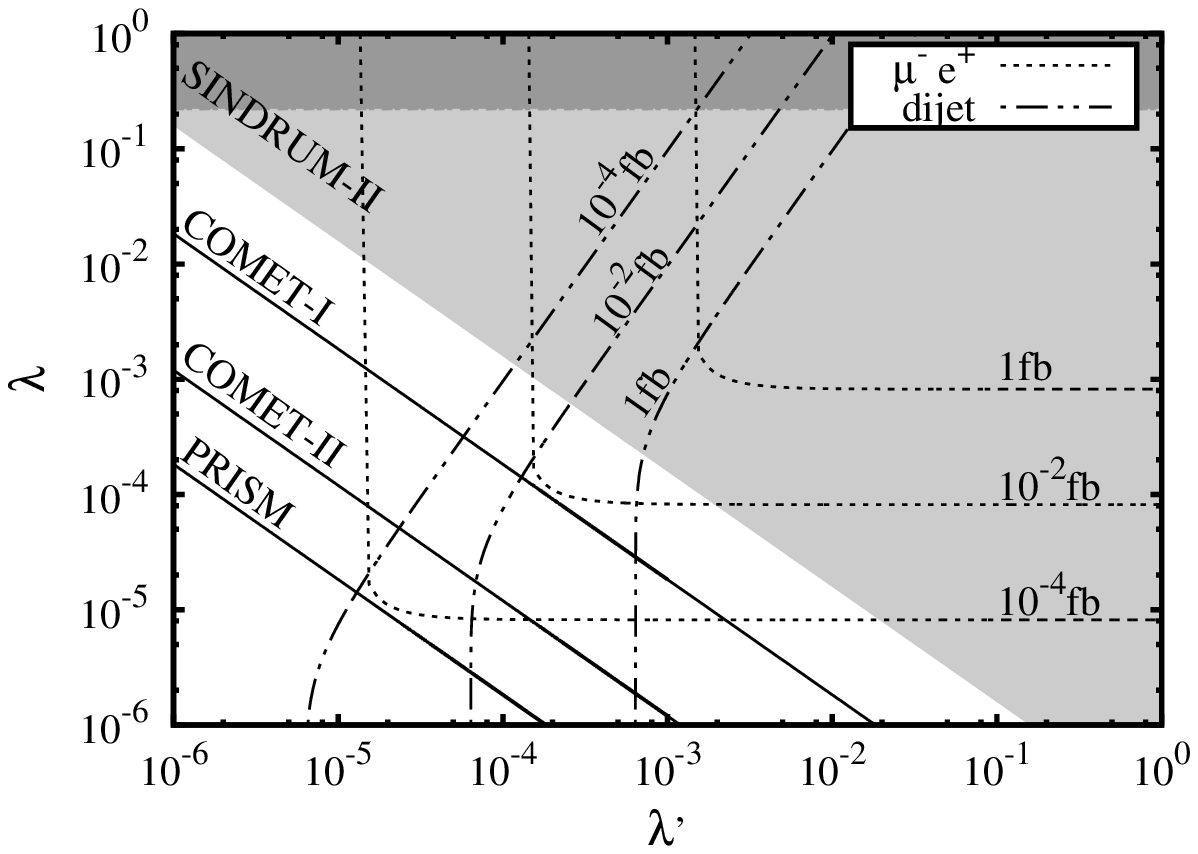}
\label{right1}
} \\
\subfigure[$m_{\tilde \nu_\tau} = 3$TeV. $\sqrt{s}= 14$TeV. ]{
\includegraphics[scale=0.61]{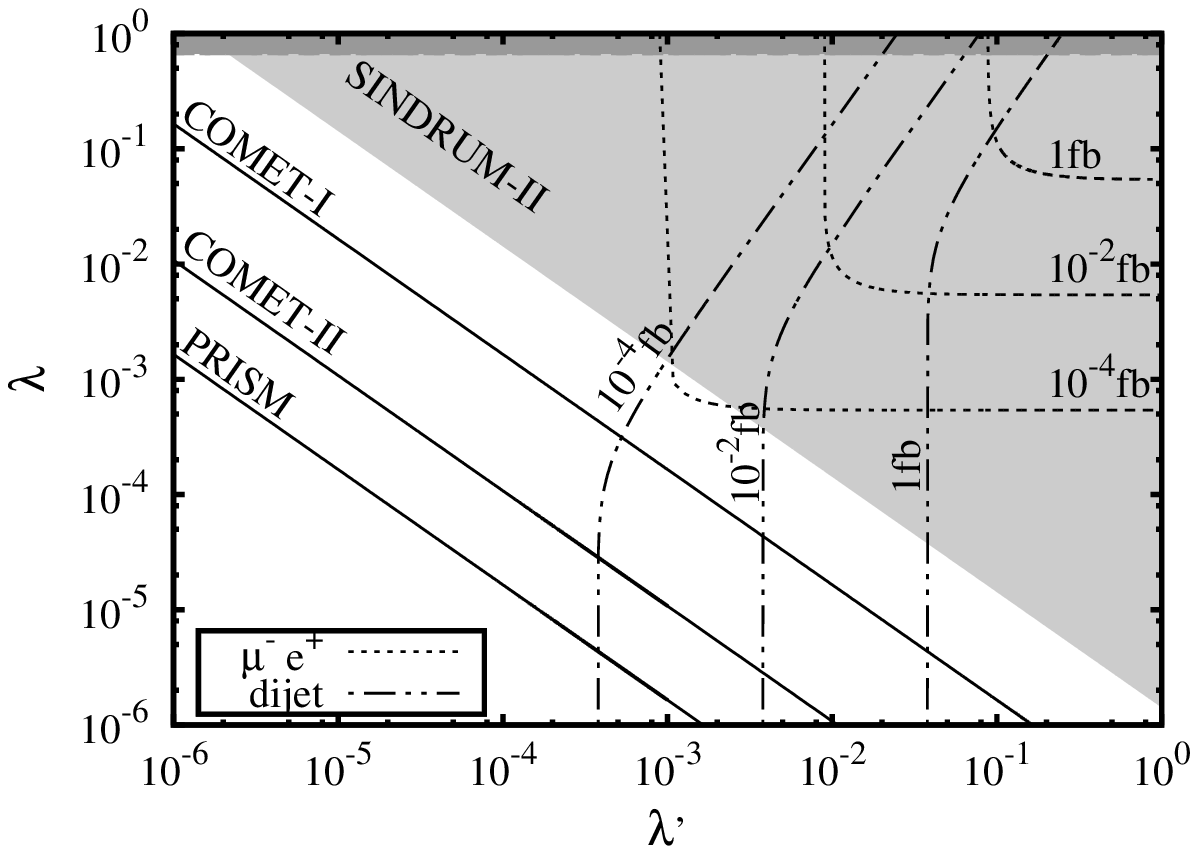}
\label{left2}
} & \hspace{-10mm}
\subfigure[$m_{\tilde \nu_\tau} = 3$TeV. $\sqrt{s}= 100$TeV. ]{
\includegraphics[scale=0.61]{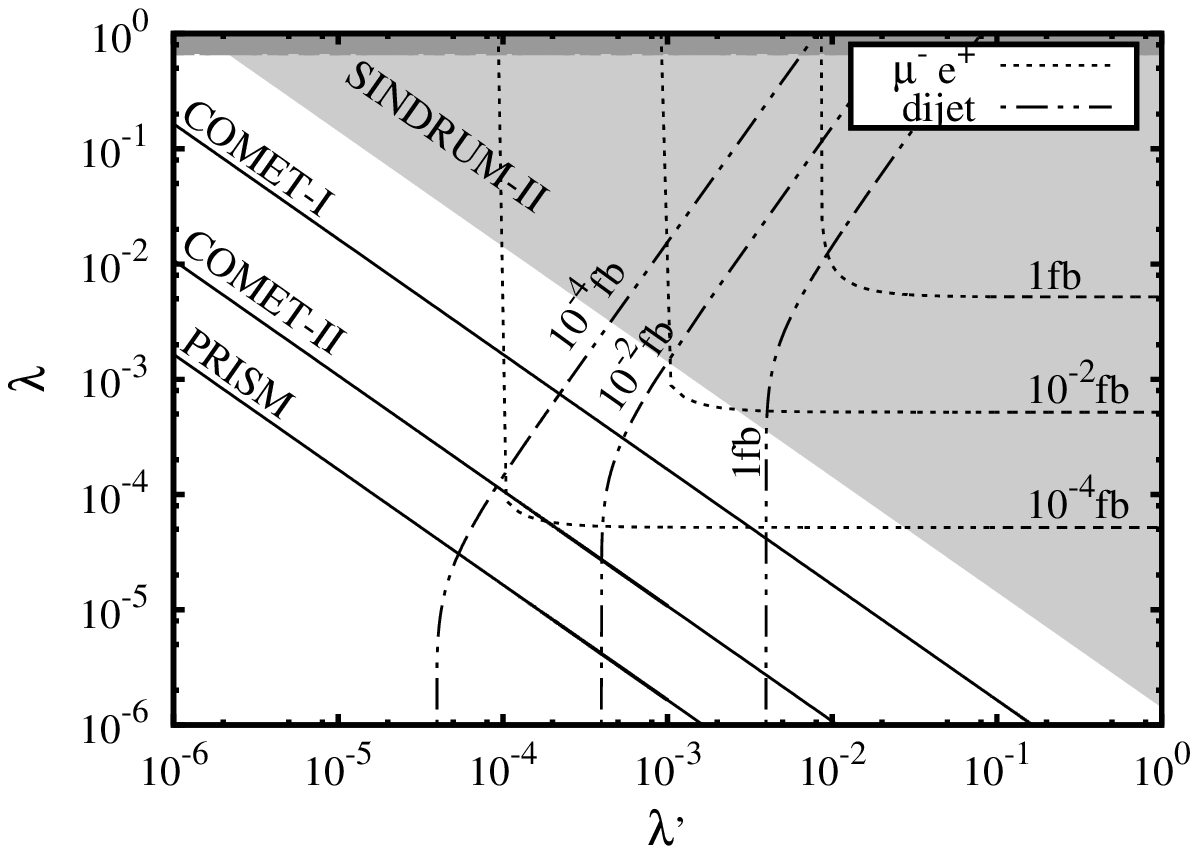}
\label{right2}
} \\
\end{tabular}
\caption{Contour plot of $\sigma(pp \to \mu^- e^+)$, $\sigma(pp 
\to dijet)$, and BR($\mu^- N \to e^- N$) in the case-I for 
(a) $m_{\tilde \nu_\tau} = 1$TeV and $\sqrt{s}=14$TeV 
(b) $m_{\tilde \nu_\tau} = 1$TeV and $\sqrt{s}=100$TeV 
(c) $m_{\tilde \nu_\tau} = 3$TeV and $\sqrt{s}=14$TeV 
(d) $m_{\tilde \nu_\tau} = 3$TeV and $\sqrt{s}=100$TeV. 
Light shaded region is excluded by the 
$\mu$-$e$ conversion search~\cite{Bertl:2006up}, 
and dark shaded band is excluded region by the $M$-$\bar M$ 
conversion search~\cite{Willmann:1998gd}.}
\label{Fig:cont_I}
\end{figure}

\begin{figure}[t!]
\hspace{-6mm}
\begin{tabular}{cc}
\subfigure[$\text{N}=\text{Si}$ and $\sqrt{s}= 14$TeV.]{
\includegraphics[scale=0.58]{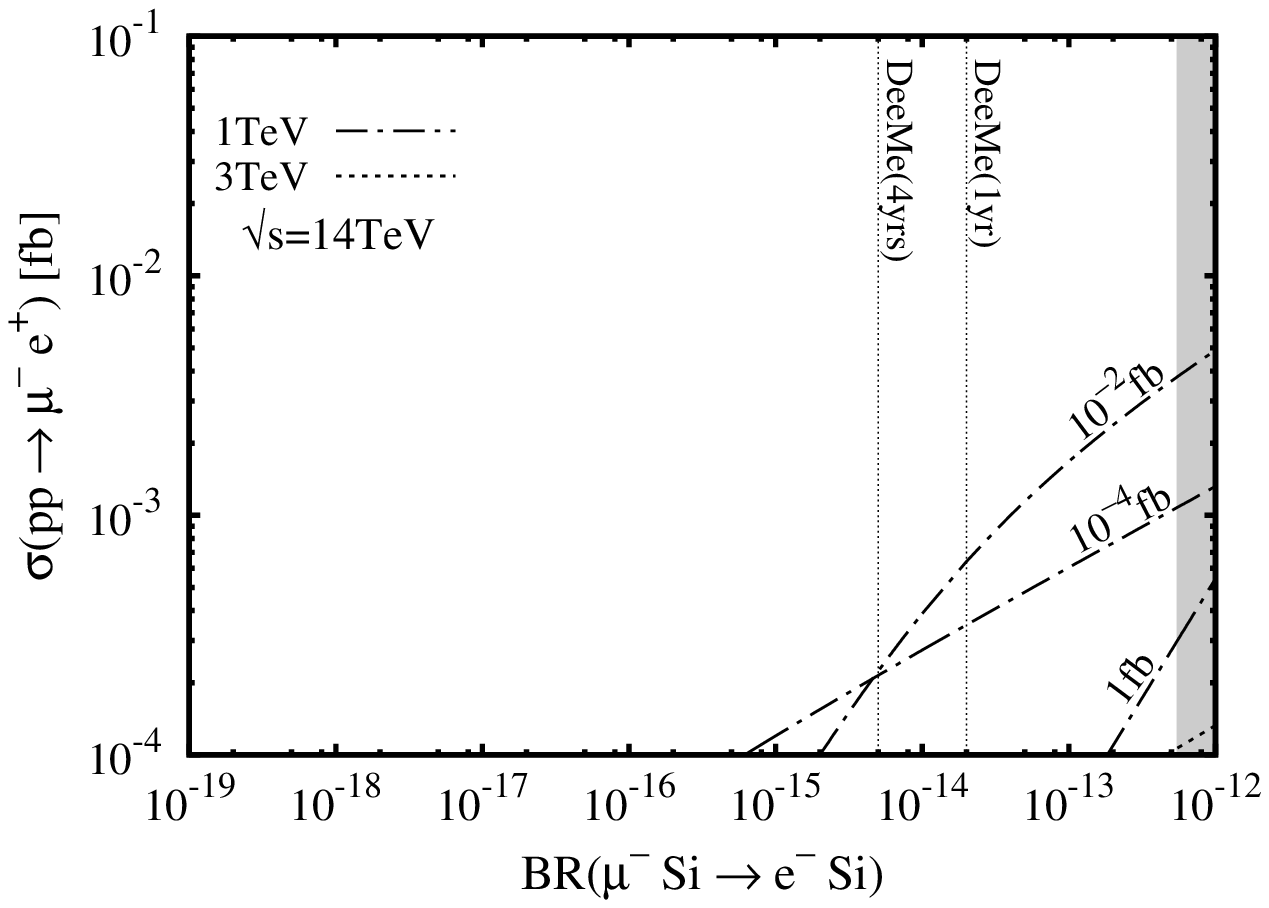}
\label{case1_a}
} & \hspace{-14mm}
\subfigure[$\text{N}=\text{Si}$ and $\sqrt{s}= 100$TeV.]{
\includegraphics[scale=0.58]{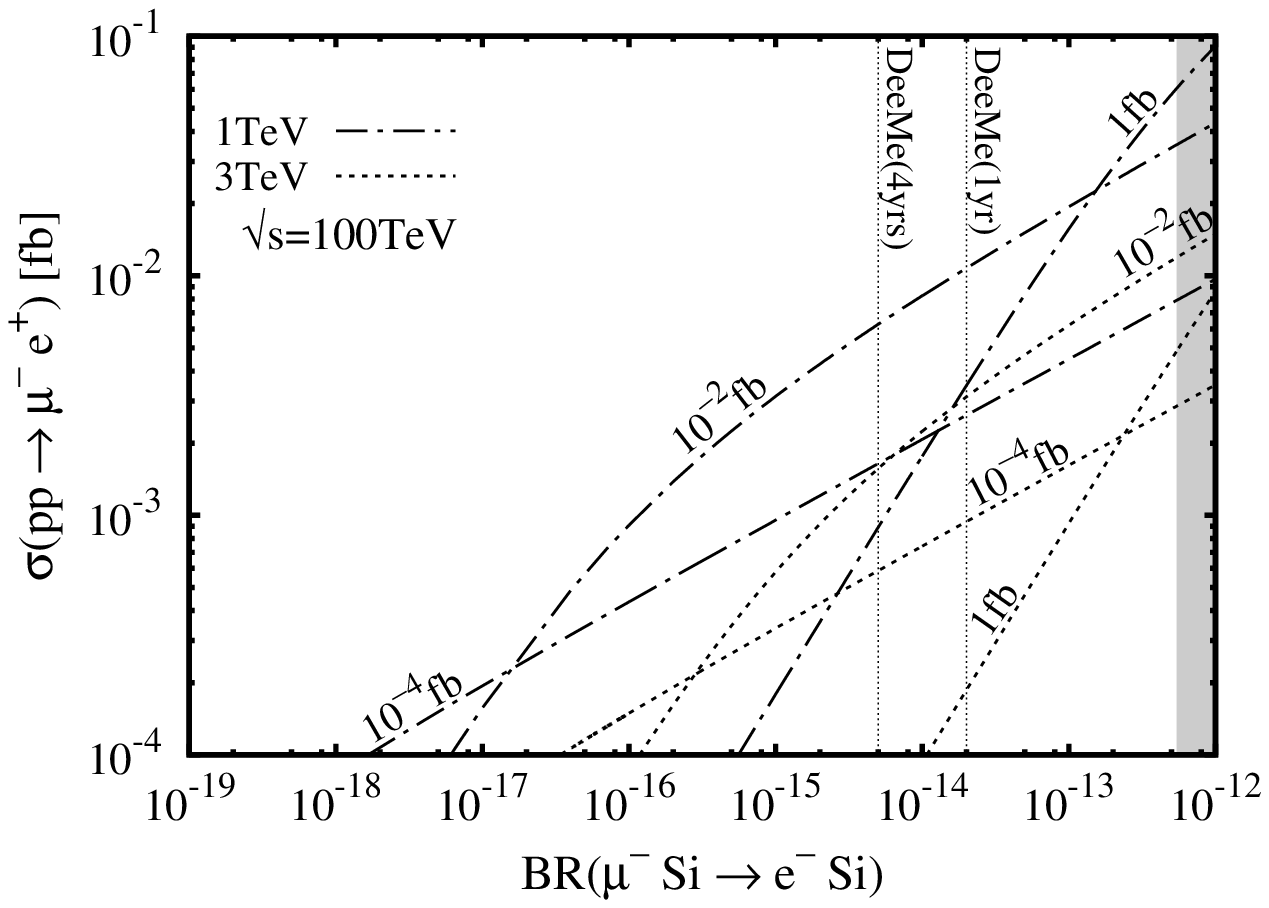}
\label{case1_b}
} \\[-4.5mm]
\subfigure[$\text{N}=\text{Al}$ and $\sqrt{s}= 14$TeV.]{
\includegraphics[scale=0.58]{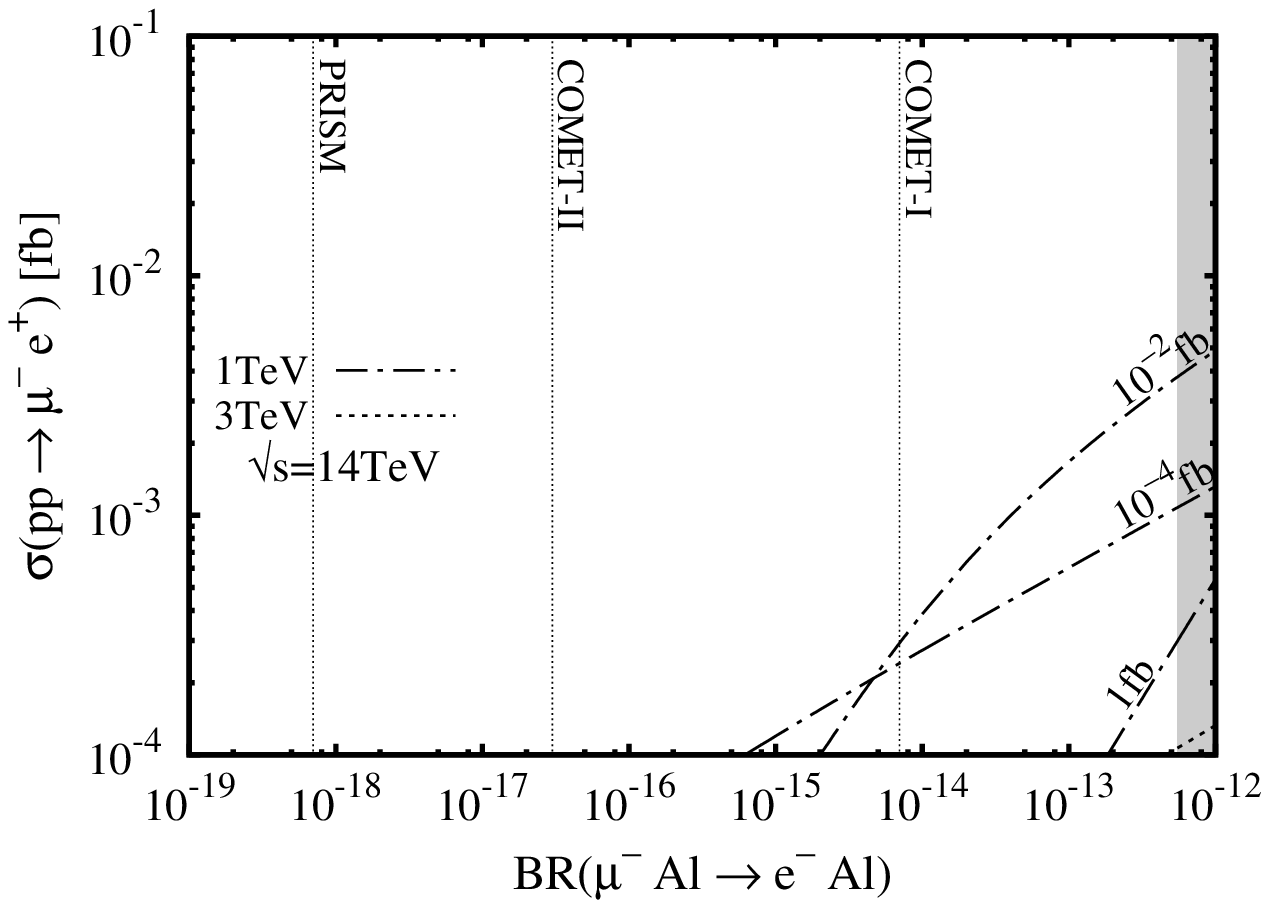}
\label{case1_c}
} & \hspace{-14mm}
\subfigure[$\text{N}=\text{Al}$ and $\sqrt{s}= 100$TeV.]{
\includegraphics[scale=0.58]{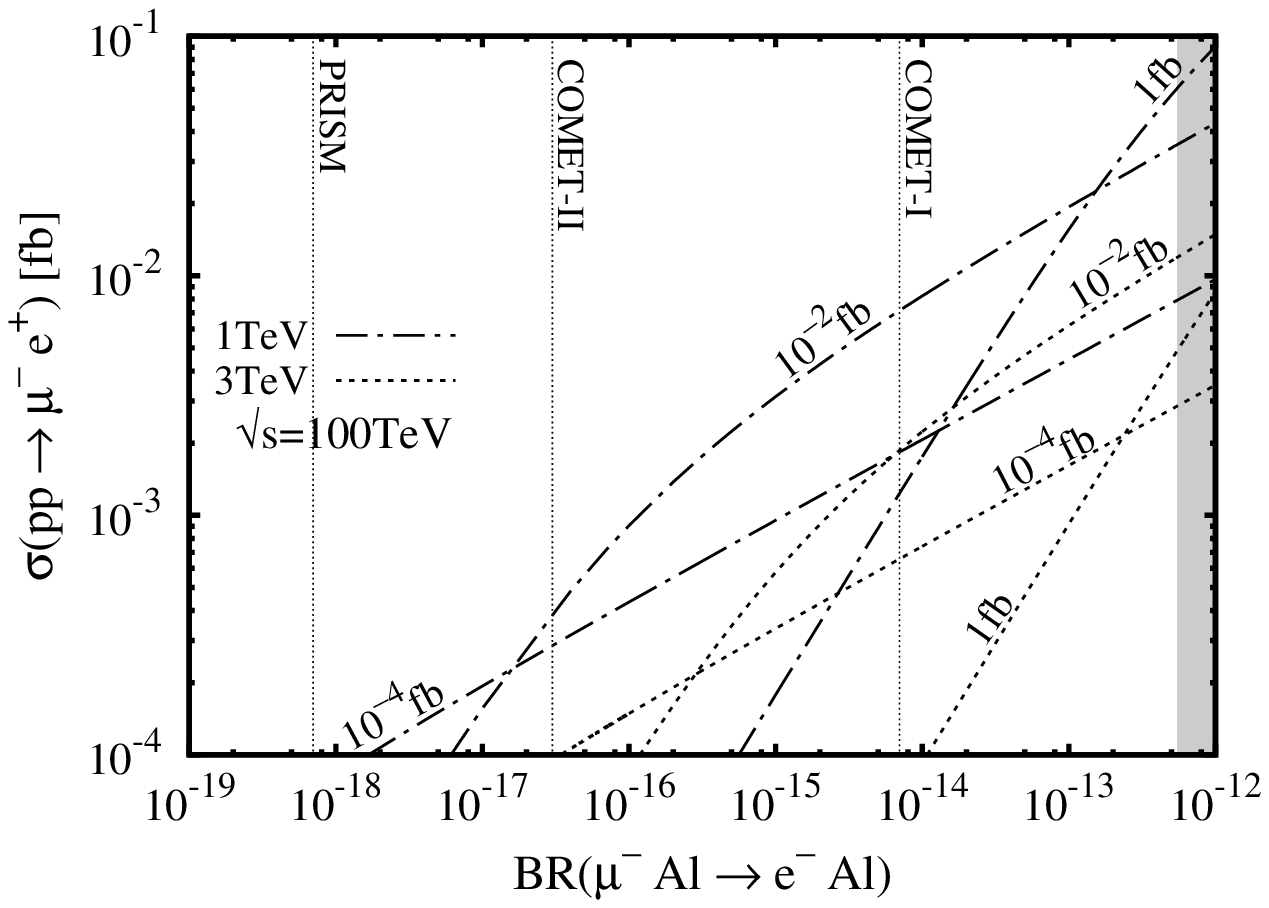}
\label{case1_d}
} \\
\end{tabular}
\caption{$\sigma (pp \to \mu^- e^+)$ as a function of $\text{BR} 
(\mu^- N \to e^- N)$ for each $\sigma (pp \to jj)$ 
in the case-I. $\sigma (pp \to jj)$ are attached on each line. 
Results for $m_{\tilde \nu_\tau} = 1\text{TeV}$ 
($m_{\tilde \nu_\tau} = 3\text{TeV}$) are given by dot-dashed 
line (dotted line). Shaded region in each panel is the excluded 
region by the SINDRUM-I\hspace{-1pt}I experiment. Left panels 
show the results 
for the collision energy $\sqrt{s} = 14\text{TeV}$, and right panels 
show the results for $\sqrt{s} = 100\text{TeV}$. We take Si [(a) 
and (b)],  and Al [(c) and (d)] for the target nucleus of 
$\mu$-$e$ conversion process. }
\label{Fig:sigma_vs_BR_I_1}
\end{figure}

The parameter dependence of $\sigma(pp \to \mu^- e^+)$, 
$\sigma(pp \to jj)$, and $\text{BR}(\mu^- N \to e^- N)$ in the case-I 
are depicted in Fig.~\ref{Fig:cont_I}. 
Dashed and dot-dashed lines are contours of $\sigma(pp \to \mu^- e^+)$ 
and $\sigma(pp \to jj)$ at $\sqrt{s}=14$TeV (left panels) and $\sqrt{s} 
=100$TeV (right panels), respectively. Solid lines are contours of 
$\text{BR}(\mu^- \text{Al} \to e^- \text{Al})$, which are translated 
from the single event sensitivities of each experiments (see 
Table~\ref{Tab:mue_conv}).  
Light shaded region is excluded by the $\mu$-$e$ conversion search at 
SINDRUM-I\hspace{-1pt}I~\cite{Bertl:2006up}, and dark shaded 
band is excluded region by the $M$-$\bar M$ conversion search 
experiment~\cite{Willmann:1998gd}. 
For simplicity, we take the couplings universally in leptonic RPV sector: 
$\lambda \equiv \lambda_{312} = \lambda_{321} = -\lambda_{132} 
= -\lambda_{231}$.

Figure~\ref{Fig:cont_I} displays the strong potential of $\mu$-$e$ 
conversion search to explore the RPV scenarios. The PRISM experiment will 
cover almost parameter space wherein the LHC experiment can survey. 
In the parameter range between the SINDRUM-I\hspace{-1pt}I limit 
and the PRISM reach, 
combining the measurement results of $\sigma(pp \to \mu^- e^+)$, 
$\sigma(pp \to jj)$, and $\text{BR}(\mu^- \text{Al} \to e^- \text{Al})$, 
the RPV couplings and the tau sneutrino mass will be precisely determined.

Figure~\ref{Fig:sigma_vs_BR_I_1} shows $\sigma(pp \to \mu \bar e)$ 
as a function of $\text{BR} (\mu + N \to e + N)$ in the case-I. 
Candidate materials for the target of $\mu$-$e$ conversion search are 
silicon (Si) at the DeeMe, and are aluminum (Al) at the 
COMET, Mu2e, and PRISM. 
Vertical dotted lines show the experimental reach of DeeMe 1-year running 
(DeeMe(1yr)), DeeMe 4-years running (DeeMe(4yrs)), COMET phase-I 
(COMET-I), COMET phase-I\hspace{-1pt}I (COMET-I\hspace{-1pt}I), 
and PRISM (PRISM). Shaded regions are the excluded region by the 
SINDRUM-I\hspace{-1pt}I~\cite{Bertl:2006up}, which 
are translated into the limit for each nucleus from that for Au. 
Each line corresponds to the dijet production cross section at the LHC, 
$\sigma(pp \to jj)$, at $\sqrt{s}=14\text{TeV}$ (left panels) and at 
$\sqrt{s}=100\text{TeV}$ (right panels), respectively. 
For simplicity, we take universal RPV coupling, $\lambda \equiv 
\lambda_{312} = \lambda_{321} = -\lambda_{132} = 
-\lambda_{231}$.

Figure~\ref{Fig:sigma_vs_BR_I_1} shows 
the clear correlations among $\sigma (pp \to \mu^- e^+)$, $\sigma(pp \to jj)$, 
and $\text{BR} (\mu^- N \to e^- N)$. Checking the correlations makes 
possible to distinguish the RPV scenario and other new physics scenarios.

\section{Summary and discussion}  \label{Sec:summary} 
\vspace{-2.5mm}

We have studied a supersymmetric standard model without R parity
as a benchmark case that COMET/DeeMe observe $\mu - e$ conversion 
prior to all the other experiments observing new physics. 

In this case with the assumption that only the third generation sleptons
contribute to such a process, we need to assume that $\{\lambda'_{311} 
{\rm \hspace{1.2mm} and/or} \hspace{1.2mm} \lambda'_{322}\} \times
\{\lambda_{312}{\rm{ \hspace{1.2mm} and/or \hspace{1.2mm} }}
\lambda_{321}\}$ must be sufficiently large. 
Though other combinations of coupling constants can lead a significant 
$\mu-e$ conversion process, only those are considered here. 
This is because in most of scenarios in the supersymmetric theory, the 
third generation of the scalar lepton has the lightest mass.

With these assumptions, we calculated the effects on future experiments.
First we considered the sensitivity of the future $\mu - e$ conversion
experiments on the couplings and the masses.

Then with the sensitivity kept into mind
we estimated the reach to the couplings by calculating the cross section
of $pp \rightarrow \mu^- e^+$ and $pp \rightarrow jj$  as a function
of the slepton masses and the couplings. To have a signal of $\mu^- e^+$
both the coupling $\lambda'$ and $\lambda$ must be large and hence
there are  lower bounds for them
while to observe dijet event via the slepton only the coupling
$\lambda'$ must be large and hence there is a lower bound on it. 
In all cases we have a chance to get confirmation of $\mu - e $
conversion in LHC indirectly. In addition, we put a bound
on the couplings by comparing both modes.

Finally we considered muonium conversion. If $\lambda'$ is very small
we cannot expect a signal from LHC. In this case at least one of $\lambda_{312}$
and $\lambda_{321}$ must be very large and if it is lucky, that is both
of them are very large we can expect muonium conversion.

There are other opportunities to check the result on $\mu -e$ conversion.
For example we can distinguish $\lambda_{312}$ and $\lambda_{321}$
in linear collider with polarized beam. We can also expect the
signal $p e^- \to p \mu^-$ in LHeC. It is however beyond the scope of this
paper to estimate their sensitivities and we leave them in future work~\cite{future}.

\section*{Acknowledgments}   
\vspace{-2.5mm}

This  work was supported in part by the Grant-in-Aid for the Ministry of Education,
Culture, Sports, Science, and Technology, Government of Japan,
No. 25003345 (M.Y.).

\vspace{-2mm}


\begin{thebibliography}{99} 

\bibitem{RPV}
  J.~Sato and M.~Yamanaka,
  arXiv:1409.1697 [hep-ph].



\bibitem{Kuno:1999jp} 
  Y.~Kuno and Y.~Okada,
  Rev.\ Mod.\ Phys.\  {\bf 73}, 151 (2001)
  [hep-ph/9909265].

\bibitem{Brooks:1999pu} 
  M.~L.~Brooks {\it et al.}  [MEGA Collaboration],
  Phys.\ Rev.\ Lett.\  {\bf 83}, 1521 (1999)
  [hep-ex/9905013].

\bibitem{Adam:2013mnn} 
  J.~Adam {\it et al.}  [MEG Collaboration],
  Phys.\ Rev.\ Lett.\  {\bf 110}, no. 20, 201801 (2013)
  [arXiv:1303.0754 [hep-ex]].

\bibitem{Bertl:2006up} 
  W.~H.~Bertl {\it et al.}  [SINDRUM II Collaboration],
  Eur.\ Phys.\ J.\ C {\bf 47}, 337 (2006).

\bibitem{Bellgardt:1987du} 
  U.~Bellgardt {\it et al.}  [SINDRUM Collaboration],
  Nucl.\ Phys.\ B {\bf 299}, 1 (1988).

\bibitem{Fukuda:1998mi} 
  Y.~Fukuda {\it et al.}  [Super-Kamiokande Collaboration],
  Phys.\ Rev.\ Lett.\  {\bf 81}, 1562 (1998)
  [hep-ex/9807003].

\bibitem{Abe:2013hdq}
  K.~Abe {\it et al.}  [T2K Collaboration],
  Phys.\ Rev.\ Lett.\  {\bf 112}, 061802 (2014)
  [arXiv:1311.4750 [hep-ex]].


\bibitem{Cui:2009zz} 
  Y.~G.~Cui {\it et al.}  [COMET Collaboration],
  KEK-2009-10.


\bibitem{Kuno:2013mha}
  Y.~Kuno [COMET Collaboration],
  PTEP {\bf 2013} (2013) 022C01.

\bibitem{Natori:2014yba} 
  H.~Natori [DeeMe Collaboration],
  Nucl.\ Phys.\ Proc.\ Suppl.\  {\bf 248-250}, 52 (2014).


\bibitem{Hisano:1995cp} 
  J.~Hisano, T.~Moroi, K.~Tobe and M.~Yamaguchi,
  Phys.\ Rev.\ D {\bf 53}, 2442 (1996)
  [hep-ph/9510309].

\bibitem{Sato:2000ff} 
  J.~Sato and K.~Tobe,
  Phys.\ Rev.\ D {\bf 63}, 116010 (2001)
  [hep-ph/0012333].

\bibitem{deGouvea:2000cf} 
  A.~de Gouvea, S.~Lola and K.~Tobe,
  Phys.\ Rev.\ D {\bf 63}, 035004 (2001)
  [hep-ph/0008085].


\bibitem{Weinberg:1981wj}
  S.~Weinberg,
  Phys.\ Rev.\ D {\bf 26} (1982) 287.


\bibitem{Sakai:1981pk}
  N.~Sakai and T.~Yanagida,
  Nucl.\ Phys.\ B {\bf 197} (1982) 533.


\bibitem{Hall:1983id}
  L.~J.~Hall and M.~Suzuki,
  Nucl.\ Phys.\ B {\bf 231} (1984) 419.

\bibitem{Koike:2010xr} 
  M.~Koike, Y.~Kuno, J.~Sato and M.~Yamanaka,
  Phys.\ Rev.\ Lett.\  {\bf 105}, 121601 (2010)
  [arXiv:1003.1578 [hep-ph]].





\bibitem{Willmann:1998gd} 
  L.~Willmann, P.~V.~Schmidt, H.~P.~Wirtz, R.~Abela, V.~Baranov, 
  J.~Bagaturia, W.~H.~Bertl and R.~Engfer {\it et al.},
  Phys.\ Rev.\ Lett.\  {\bf 82}, 49 (1999)
  [hep-ex/9807011].



\bibitem{future}
 J.~Sato and M.~Yamanaka, in preparation. 





\end{thebibliography}
\end{document}